\documentstyle[pra,aps,epsfig]{revtex}
\begin{document}
\draft

\title{Quantum statistics of atoms in microstructures}
\author{Erika Andersson, M\'arcia T. Fontenelle and Stig Stenholm}
\address{Department of Physics,
Royal Institute of Technology, Lindstedtsv. 24,
SE-10044 Stockholm, Sweden}
\date{\today}
\maketitle

\begin{abstract}
This paper proposes groove-like potential structures for the
observation of quantum information processing by trapped particles. As an
illustration the effect of quantum statistics at a 50-50 beam splitter
is investigated. For non-interacting particles we regain the results
known from photon experiments, but we have found that particle interactions
destroy the perfect bosonic correlations. Fermions avoid each other due
to the exclusion principle and hence they are far less sensitive to
particle interactions. For bosons, the behavior can be explained with
simple analytic considerations which predict a certain amount of
universality. This is verified by detailed numerical calculations.
\end{abstract}

\section{Introduction}

The duality between wave and particle aspects is one of the central issues
of Quantum Mechanics. Much has been made of the particle aspects of photons,
but the recent progress in cooling and controlling atomic motion has brought
forward their wave mechanical behavior in a prominent way. The new field of
Atomic Optics has emerged \cite{ada}. 

With modern cooling and trapping techniques, one can envisage controlled 
motion of atomic particles in structures whose mechanical dimensions
match the heterostructures used in electronic circuits. Neutral atoms can be
stored in magneto-optical traps, and H\"{a}nsch and his group
has recently shown \cite{haen} that such traps can be made very small,
{\it viz} of the order of $10^2\mu{\rm m}.$
This requires high precision in the fabrication of the solid structures
defining the dimensions of the trap. Modern lithographic technology
suggests that such structures could be made even much smaller, and then we
can imagine experiments in traps of genuinely microscopic dimensions, where
quantum effects would dominate the particle dynamics. H\"{a}nsch has also
suggested that such traps could be made into channels and structures, thus
providing a tool to design arbitrary devices at the surface of a substrate.

Similar structures can be constructed by combining charged wires with
evanescent wave mirrors \cite{ess,sei} or magnetic mirrors \cite{roa}.
Such combinations can be used to build up the structures utilized in
nano-electronics. The use of wires to guide atomic motion has been
investigated by Denschlag and Schmiedmayer \cite{den}. Schmiedmayer
has also discussed the use of such structures to construct quantum
dots and quantum wires for atoms \cite{sch}. 

Alternative ways to achieve guided motion and possibly controlled
interaction between atoms is to utilize hollow optical fibers with
evanescent waves trapping the atoms to narrow channels at the center of
the fiber \cite{ito,yin}. These can eventually be fused to provide
couplers similar to those used for optical signal transmission in
fibers. Also the pure atomic waveguide achievable by the use of hollow
laser modes may be used.

We see these methods as an opening to novel and innovative uses 
of particle traps. By arranging a network of grooves on a surface, 
we can launch particles (wave packets) into the various inputs of 
the system, let them propagate through the device and interact with 
its structures and each other. This may well provide an opportunity 
to design quantum apparatuses, process information and perform 
computations. The advantage is that both the structures and the input 
states are easy to control in an atomic environment. An equivalent
point of view is expressed by Schmiedmayer in Ref. \cite{sch}.

A next step in the experimental progress would be to observe the quantum 
character of atoms (or possibly ions).
An essential quantum characteristic of particles is their statistical
behavior. The difference between bosons and fermions manifests
itself dramatically in many  situations. Optical networks can be
fed by a few photons only, and their quantum aspects have been
utilized in experiments
ranging from secure communication to tests of fundamental issues. Recently
Zeilinger and his group \cite{zeil} have tested the behavior of two-photon
states at beam splitters. Using the overall symmetry properties of the
states, they have been able to display both symmetric and antisymmetric
behavior. 

Similar experiments are in principle possible with electrons. In
nanostructures, one can fabricate the devices simulating optical components,
but it is far less trivial to launch single conduction electrons in well
controlled states. Yamamoto's group, however, has been able to show quantum
correlations in an experiment which is the analogue of a beam splitter  for
photons \cite{yama}.

In this paper we give an example of the multiparticle effects
observable when particle states are launched along potential grooves
on a surface. The specific phenomenon singled out for investigation is
the effect of particle statistics at a beam-splitter-like coupling
device. The corresponding effect with photons is described in Sec. II
as a motivation.
In Sec. III we present the details of the model chosen and a
simplified analytic treatment demonstrating the main features expected
of this model. In Sec. IV we carry through a numerical analysis of the
situation, for one particle as a two-dimensional propagation problem,
but for two particles in a paraxial approximation. For non-interacting
particles, the expected behavior is found, but when particle
interactions are added, the boson behavior is changed. For fermions
the exclusion principle make them essentially insensitive to the
interaction. An unexpected feature is found: the sign of the
interaction is irrelevant for the effect. In Sec. V this is explained
within our simple analytic model, and, as a consequence, a certain
universality is proposed: When the interaction strength over the
tunneling frequency becomes of the order of $\sqrt{3}$, the
noninteracting bosonic behavior is essentialy destroyed. This is
verified by numerical calculations, reported in Fig. 13. Finally
Sec. VI presents a discussion of parameter ranges in real materials,
where our effects may be observable, and summarizes our conclusions. 

\section{Motivation}

In order to show the opportunities offered by atomic networks, we 
investigate the
manifestations of quantum statistics on an experiment emulating the
behavior of photons in beam splitters. This is a straightforward approach,
which enables us to display the potentialities and limitations of such
treatments.

\begin{figure}
\centerline{\epsffile{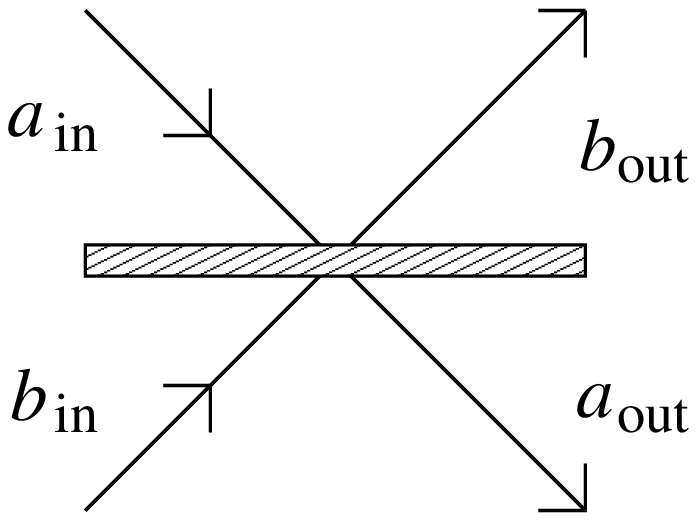}}
\caption{Schematic drawing of a beam splitter. The incoming
modes $a_{\rm{in}}, b_{\rm{in}}$ are piloted into the outgoing modes 
$a_{\rm{out}}, b_{\rm{out}}$ according to the beam splitter relations.}
\label{bs}
\end{figure}

Our work has been motivated by the statistics displayed by a 50-50
beam splitter, which has been used in the experiments by the Zeilinger
group \cite{zeil}.
When two particles are directed into the beam splitter in the incoming
modes  in Fig. \ref{bs},
they are piloted into the outgoing modes according to the beam splitter
relations 
\begin{equation}
\left[\begin{array}{c}
a_{\rm{out}}^{\dagger } \\ 
b_{\rm{out}}^{\dagger }
\end{array}\right]
=\frac 1{\sqrt{2}}
\left[\begin{array}{c c}
1 & -i \\ 
-i & 1
\end{array}\right]
\left[\begin{array}{c}
a_{\rm{in}}^{\dagger } \\ 
b_{\rm{in}}^{\dagger }
\end{array}\right];  \label{a1}
\end{equation}
see Ref. \cite{leo}.
When one particle is directed into each incoming channel, the state is 
\begin{equation}
| \Psi \rangle =a_{\rm{in}}^{\dagger }b_{\rm{in}}^{\dagger }| 0\rangle ,
\label{a2}
\end{equation}
where $|$$0\rangle$ is the vacuum state. Without assuming anything about
the statistics of the incoming particles, we can express the state 
(\ref{a2}) in terms of the outgoing states by inverting the relation 
(\ref{a1}) as 
\begin{equation}
| \Psi \rangle =\frac i2\left[ \left( a_{\rm{out}}^{\dagger }\right)
  ^2+\left(
b_{\rm{out}}^{\dagger }\right) ^2\right] | 0\rangle +\frac 12\left[
a_{\rm{out}}^{\dagger },b_{\rm{out}}^{\dagger }\right] | 0\rangle .  \label{a3}
\end{equation}
From this follows that boson statistics gives 
\begin{equation}
| \Psi \rangle =\frac i{\sqrt{2}}\left( | n_{a,\rm{out}}=2,n_{b,\rm{out}}=
0\rangle +| n_{a,\rm{out}}=0,n_{b,\rm{out}}=2\rangle \right) ;  \label{a4}
\end{equation}
the particles emerge together at either output. For fermions we have 
\begin{eqnarray}
| \Psi \rangle &=&a_{\rm{out}}^{\dagger }b_{\rm{out}}^{\dagger }
| 0\rangle \label{a5}\\
&=& | n_{a,\rm{out}}=1,n_{b,\rm{out}}=1\rangle, \nonumber
\end{eqnarray}
and they always remain separated.

Weihs et al. \cite{zeil} have been able to verify these properties
experimentally using photons. As the requirements of quantum statistics
refer only to the total wave functions, they have been able to realize both
the symmetric and the antisymmetric case, thus offering the behavior of
both bosons and fermions.

Photons are ideal for experiments, they do not interact mutually and they
propagate essentially undisturbed in vacuum. As models for quantum systems,
they have the drawback that they cannot be localized, their wave packets are
of rather elusive character, and the influence of particle interactions
cannot be established. Thus we have chosen to discuss the propagation of
massive particles through beam-splitter-like structures. As explained above,
such experiments may be performed with atoms or electrons in traps of
microscopic dimensions. We can thus investigate the propagation of wave
packets through these structures, explore the role of quantum statistics
and switch on and off the particle interaction at will.

\section{The model}
\label{smod}

We consider particles moving in potential
wells which form grooves over a two-dimensional surface. These may cross or
couple by tunneling when approaching each other, thus forming a network of
potential channels emulating a linear optical system.

\begin{figure}
\centerline{\epsffile{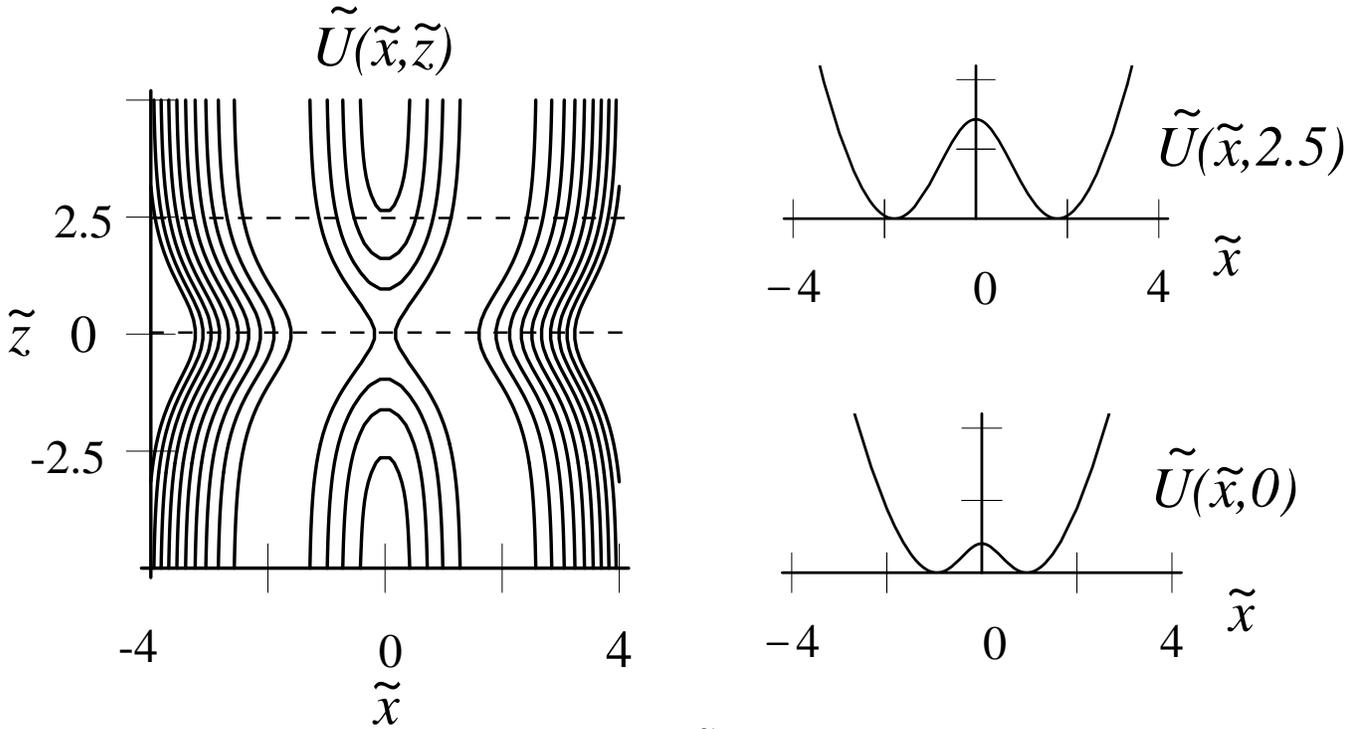}}
\caption{A contour plot of the beam splitter potential 
$\widetilde{U}(\widetilde{x}
,\widetilde{z})$. The two channels approach each other at 
$\widetilde{z}=0$. The scaled variables $\widetilde{x}$ and 
$\widetilde{z}$ used in the numerical calculations are related 
to $x$ and $z$ according to Eq. (\ref{a20}). The scaled dimensionless 
oscillator frequency is $\widetilde{\omega}=30$. The distance 
$d(\widetilde{z})$ between the two valleys is chosen as in Eq. 
(\ref{a27}), with $d_0=1.8903$ and $\eta=1$. Two cross sections of
the potential at $\widetilde{x}=0$ and $\widetilde{x}=2.5$ are also shown.}
\label{pot}
\end{figure}

Here we consider two separate channels which run parallel for $z\rightarrow
\pm \infty $ and approach each other in the $x-$direction, as shown in
Fig. \ref{pot}. For simplicity we construct the potential from two harmonic
oscillators 
\begin{equation}
U_{\pm }(x)=\frac 12m\omega ^2(x\pm{1\over 2}d)^2,  \label{a6}
\end{equation}
so that a double well potential can be obtained by
writing 

\begin{eqnarray}
U(x,z) & = & {U_{+}(x,z)U_{-}(x,z)}\over{U_{+}(x,z)+U_{-}(x,z)}\label{a7}\\  
& = & \frac 12m\omega ^2 {{\left( x+{1\over 2}d(z)\right) ^2
\left( x-{1\over 2}d(z)\right)  ^2}
\over{\left( x+{1\over 2}d(z)\right) ^2+\left( x-{1\over 2}d(z)\right)
  ^2}} \nonumber.
\end{eqnarray}

If we now choose $d(z)$ in a suitable manner, we can achieve the potential
behavior shown in Fig. \ref{pot}. Note that at the minima, the potential 
$U(x,z)$ essentially follows the shape of the smaller potential $U_{\pm }$.

We consider a wave packet sitting stationary near the bottom of one
well at 
$z=0,$ where the distance between the wells is at its minimum $d_0$. 
The particle can then tunnel across the barrier with the rate 
\begin{equation}
T\sim \exp \left[ -\int \sqrt{2mU(x,0)}dx\right] \approx 
\exp \left[ -\kappa \sqrt{U(0,0)}d_0\right] ,  \label{a8}
\end{equation}
where $\kappa $ is some constant. From Eq. (\ref{a7}) we see that 
$U(0,0)\propto d_0^2$ so that we expect 
\begin{equation}
\log T\sim -\kappa ^{\prime }d_0^2+\rm{const}.  \label{a9}
\end{equation}

\begin{figure}
\centerline{\epsffile{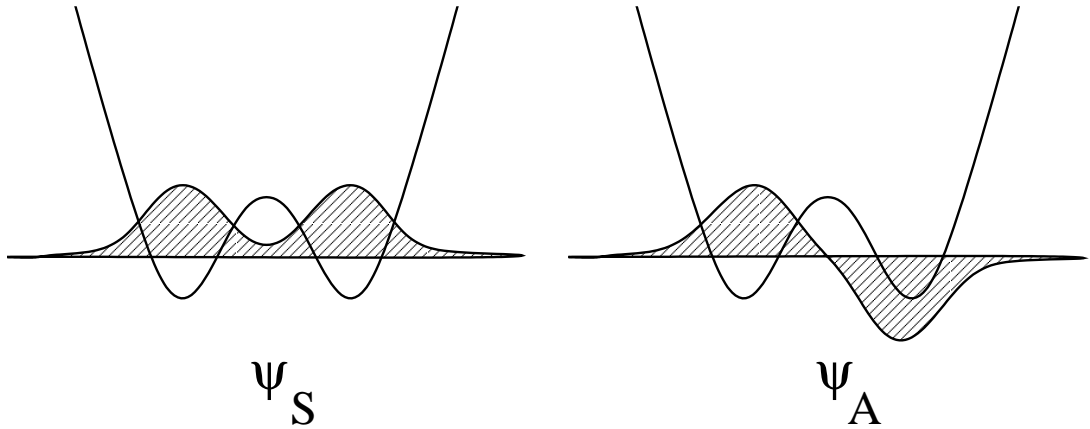}}
\caption{Symmetric and antisymmetric eigenfunctions $\psi_S$ and $\psi_A$ 
of the double well. }
\label{eigen}
\end{figure}

In order to acquire a heuristic understanding of the physics involved in the
coupling of the grooves at $z=0,$ we look at the lowest eigenfunctions
of the double well potential. These are expected to be symmetric, 
$\psi _S$, with energy $E_S$, and antisymmetric, $\psi _A$, with
energy $E_A$, as shown in Fig. \ref{eigen}.
We have $E_A>E_S$ and hence we write 
\begin{eqnarray}
E_A & = & \overline{E}+\hbar \Omega  \nonumber\\ 
E_S & = & \overline{E}-\hbar \Omega,
\label{a11}
\end{eqnarray}
where $2\Omega $ is the tunneling frequency.

Using the eigenstates we form the localized states 
\begin{eqnarray}
\varphi _L & = & \frac 1{\sqrt{2}}\left( \psi _S+\psi _A\right) \nonumber\\ 
\varphi _R & = & \frac 1{\sqrt{2}}\left( \psi _S-\psi _A\right),
\label{a12}
\end{eqnarray}
where the subscripts $L$ $(R)$ denote left (right) localization.

We can easily integrate the time evolution by using the energy 
eigenstates. If we now assume that we start from $\varphi _L$ at time 
$t=0$, then 
\begin{eqnarray}
\Psi (t) & = & \exp \left( -iHt/\hbar \right) \varphi _L \nonumber \\ 
& = & {1\over\sqrt{2}} \exp \left( -i\overline{E}t/\hbar \right)\left(
e^{i\Omega t}\psi _S+e^{-i\Omega t}\psi _A\right)  \label{a13}\\ 
& = & \exp \left( -i\overline{E}t/\hbar \right) \left( \cos \Omega
t\,\varphi _L+i\sin \Omega t\,\varphi _R\right) .\nonumber
\end{eqnarray}
This displays the expected flipping back and forth between the two wells.
For 
\begin{equation}
\Omega t_0=\frac \pi 4  \label{a14}
\end{equation}
the coupling performs the action of a 50-50 beam splitter.

We now move to consider the action of such a potential configuration on a
two particle initial state. We first choose the bosonic one 
\begin{equation}
\Psi _0^B=\frac 1{\sqrt{2}}\left( \varphi _L(1)\varphi _R(2)+\varphi
_L(2)\varphi _R(1)\right) ,  \label{a15}
\end{equation}
where the argument denotes the coordinates of the particle. This can be
expressed as 
\begin{equation}
\Psi _0^B=\frac 1{\sqrt{2}}\left( \psi _S(1)\psi _S(2)-\psi _A(1)\psi
_A(2)\right) ,  \label{a16}
\end{equation}
which can be evolved in time straightforwardly to give 
\begin{eqnarray}
\exp \left( -iHt_0/\hbar \right) \Psi _0^B & = &{1\over \sqrt{2}} 
\exp \left( -i2\overline{E}t_0/\hbar \right) \left( e^{i2\Omega t_0}\psi
_S(1)\psi _S(2)-e^{-i2\Omega t_0}\psi _A(1)\psi _A(2)\right) \nonumber\\ 
& = & {i\over\sqrt{2}}\exp \left( -i2\overline{E}t_0/\hbar \right)
\left( \varphi _L(1)\varphi _L(2)+\varphi _R(2)\varphi _R(1)\right) .
\label{a17}
\end{eqnarray}
As we see, the bosonic two particle 
state works as in Eq. (\ref{a4}): both particles emerge together.

In the fermionic case we have  
\begin{eqnarray}
\Psi _0^F & = & \frac 1{\sqrt{2}}\left( \varphi _L(1)\varphi _R(2)-\varphi
_L(2)\varphi _R(1)\right)\nonumber\\ 
& = & \frac 1{\sqrt{2}}\left( \psi _A(1)\psi _S(2)-\psi _S(1)\psi
_A(2)\right) . \label{a18}
\end{eqnarray}
Because both states $\psi _A\psi _S$ and $\psi_S\psi_A$
evolve with the energy $2\overline{E}$, $\Psi _0^F$
remains uncoupled to other states. Thus the fermions emerge at
separate exit channels as expected.

\section{Numerical work}

\subsection{The Schr\"{o}dinger equation}
\label{num}

The Schr\"{o}dinger equation in the two-dimensional system is of the form 
\begin{equation}
i\hbar \frac \partial {\partial t}\Psi (x,z,t)=\left[ -\frac{\hbar ^2}{2m
}\left( \frac{\partial ^2}{\partial x^2}+\frac{\partial ^2}{\partial z^2}
\right) +U(x,z)\right] \Psi (x,z,t).  \label{a19}
\end{equation}
As a preparation for the numerical work, we introduce the scaling parameters 
$\tau $ and $\xi $ giving the dimensionless variables 
\begin{eqnarray}
\widetilde{x} & = & x/\xi  \nonumber\\ 
\widetilde{z} & = & z/\xi \label{a20} \\ 
\widetilde{t} & = & t/\tau \nonumber \\ 
\widetilde{p} & = & \tau p/m\xi .\nonumber
\end{eqnarray}
We apply this to the one-dimensional oscillator Hamiltonian 
\begin{equation}
H=\frac{p^2}{2m}+\frac 12m\omega ^2x^2  \label{a21}
\end{equation}
and find the Schr\"{o}dinger equation 
\begin{equation}
i\left( \frac{\hbar \tau }{m\xi ^2}\right) \frac \partial {\partial 
\widetilde{t}}\Psi =\left( \frac{\widetilde{p}^2}2+\frac 12\widetilde{\omega 
}^2\widetilde{x}^2\right) \Psi .  \label{a22}
\end{equation}
The dimensionless oscillator frequency is given by 
$\widetilde{\omega }=\omega \tau $. This shows that choosing the 
scaling units suitably, we can
tune the effective dimensionless Planck constant 
\begin{equation}
\widetilde{\hbar }=\frac{\hbar \tau }{m\xi ^2}.  \label{a23}
\end{equation}
To check the consistency of this we calculate 
\begin{equation}
\left[ \widetilde{x},\widetilde{p}\right] =\left( \frac 1\xi \right) \left(
\frac \tau {m\xi }\right) \left[ x,p\right] =i\widetilde{\hbar }.
\label{a24}
\end{equation}
This gives us a way of controlling the quantum effects in the numerical
calculations.

In our numerical calculations we employ the split operator method \cite{split}
\begin{equation}
\exp \left[ -i(T+U)\Delta t/\hbar \right] \approx 
\exp \left[ -iT\Delta t/\hbar \right] 
\exp \left[ -iU\Delta t/\hbar \right] .  \label{a25}
\end{equation}
The corrections to this are given by 
\begin{equation}
\left[ T,U\right] \frac{\Delta t^2}{2\hbar ^2}=
\left|\left( \frac{\widetilde{\Delta t}^2\widetilde{\omega}^2}4\right)
\left(\frac{\widetilde{x}\,\widetilde{p}+\widetilde{p}\,\widetilde{x}}
{\widetilde{\hbar }}\right)\right| .
\label{a26}
\end{equation}
In order to achieve satisfactory numerical accuracy, this should not be too
large; in our calculations, with $\Delta t=0.001$,
$\widetilde{\omega}=30$ and $\widetilde{\hbar}=6$
 the expectation value of expression (\ref{a26}) is of the order of $10^{-4}$.
Decreasing $\Delta t$ or the grid spacing has been found not to change
our results significantly.

In the following discussion, we use the scaled variables, but for
simplicity, we do not indicate this in the notation. Whenever variables
are assigned dimensionless values, these refer to the scaled versions.

\begin{figure}
\centerline{\epsfig{file=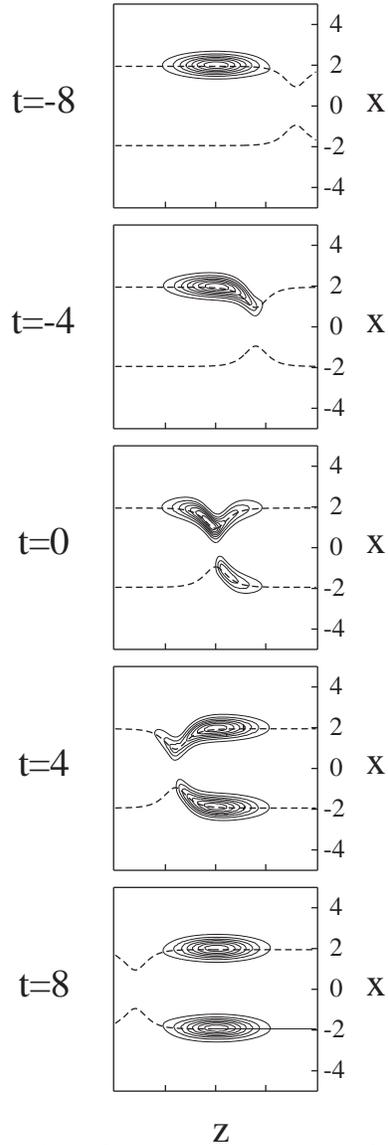,width=5cm}}
\caption{
Time evolution of a two-dimensional Gaussian wave packet
propagating through the 50-50 beam splitter. The frame shown is
centered around the wave packet moving in the
$z$-direction. The dashed lines represent the minima of the potential.
The potential is given by the scaled version of Eq.
(\ref{a7}) with $d(z)$ as in Eq. (\ref{a27}) where $\eta=30$. The
scaled oscillator frequency is $\widetilde{\omega}=30$;
$\widetilde{\hbar}=6$.} 
\label{2dwave}
\end{figure}

In order to achieve beam splitter operation, we let the distance between the
potential wells vary in the following way 
\begin{equation}
d(z)=2+d_0-\frac 2{\cosh (z/\eta)},  \label{a27}
\end{equation}
which inserted into Eq. (\ref{a7}) gives a potential surface as shown in Fig.
\ref{pot}. To test its operation as beam splitter, 
we let a wave packet approach the
coupling region in one of the channels, and follow its progress through the
intersection numerically as a two-dimensional problem. The result is shown 
in Fig. \ref{2dwave}. We see that the parameters chosen lead to ideal
50-50 splitting of
the incoming wave packet. The progress of the wave packet through the
interaction region is steady and nearly uniform, and no backscattering is
observed. This suggest simplifying the situation so that the motion in the
$z$-direction is replaced by a constant velocity, and the full quantum problem
is computed only in the $x$-direction. If the wave packet is long enough in
the $z$-direction, its velocity is well defined, and this should be a good
approximation.

The implementation of such a paraxial approach becomes imperative when we
want to put two particles into the structure. The full two-dimensional
integration would require the treatment of four degrees of freedom, which is
demanding on the computer resources. With the paraxial approximation, 
two particles  can be treated by a two-dimensional numerical approach,
which is within the resources available.

To introduce the paraxial approximation, we perform a Galilean
transformation of the wave function to a co-moving frame 
\begin{equation}
\psi (x,z,t)=\varphi (x,\varsigma ,t)\exp \left[ \frac i{\hbar }\left(
p_0z-\frac{p_0^2t}{2m}\right) \right] ,  \label{a28}
\end{equation}
where 
\begin{equation}
\varsigma =z-\frac{tp_0}m.  \label{a29}
\end{equation}
The initial momentum in the $z$-direction is denoted by $p_0$. The new wave
function is found to obey the Schr\"{o}dinger equation 
\begin{equation}
i\hbar \frac \partial {\partial t}\varphi (x,\varsigma ,t)=\left[ -\frac{
\hbar ^2}{2m}\left( \frac{\partial ^2}{\partial x^2}+\frac{\partial ^2}{
\partial \varsigma ^2}\right) +U\left( x,\varsigma +\frac{tp_0}m\right)
\right] \varphi (x,\varsigma ,t).  \label{a30}
\end{equation}
For a well defined momentum $p_0$, the wave packet is very broad in the
$\varsigma $-direction and its derivatives with respect to $\varsigma $ may
be neglected. The corresponding degree of freedom disappears, and it is
replaced by a potential sweeping by with velocity $p_0/m.$ This is what we
call the paraxial approximation.

\begin{figure}
\centerline{\epsffile{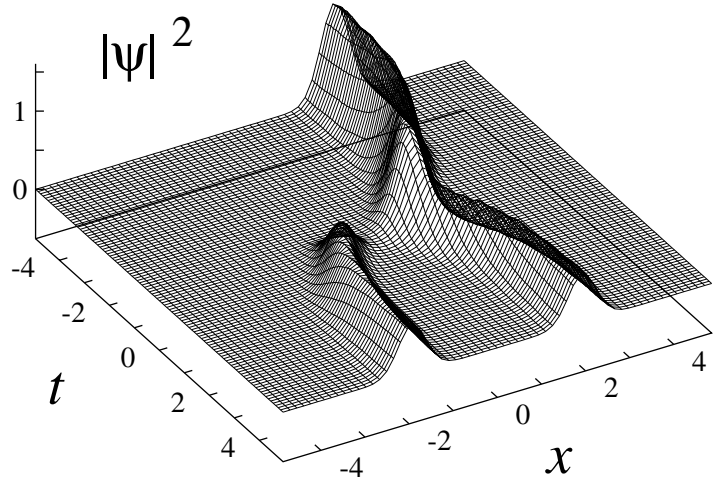}}
\caption{Time evolution of a wave packet propagating through the 50-50
beam splitter in the paraxial approximation.
The wave packet is incident in one channel and splits into two equal
parts. Parameters are as in Fig. \ref{2dwave}, the only difference is
that in the potential $U(x,z)$, $z$ is replaced by $tp_0/m$ according 
to the paraxial approximation. The initial wave packet is
$\varphi^0(x)=N\exp \left[ -{\omega\over{2\hbar}}
\left( x- (1+d_0/2)\right)^2 \right]$.}
\label{1dwave}
\end{figure}

\begin{figure}
\centerline{\epsffile{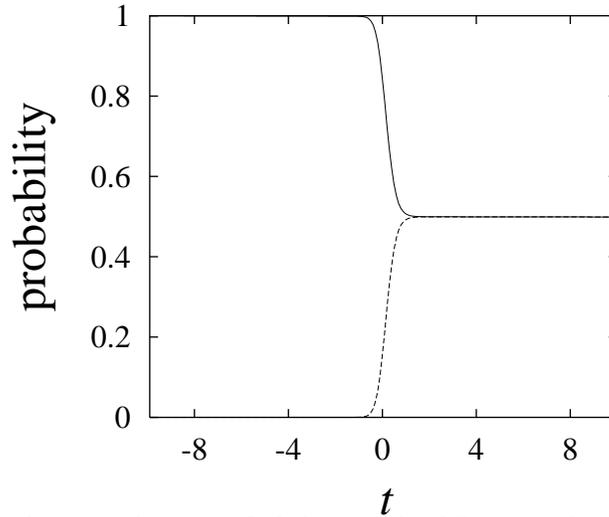}}
\caption{The probabilities, as a function of time, to find the
  particle of Fig. \ref{1dwave} in
the right (full line) and left (dashed line) valley of the beam splitter.
The particle is incident at the right input and emerges with equal
probability at the left and right outputs. Parameters as in Fig. \ref{1dwave}.}
\label{probsplit}
\end{figure}

Taking the parameters from the integration in Fig. \ref{2dwave}, we
can obtain the beam splitting operation also in the paraxial 
approximation as shown in Fig. \ref{1dwave}.
The transfer of the wave packet from one well to the linear superposition is
shown in Fig. \ref{probsplit}. This proves that the potential 
configuration works exactly as in the analytic result (\ref{a13}).

\begin{figure}
\centerline{\epsffile{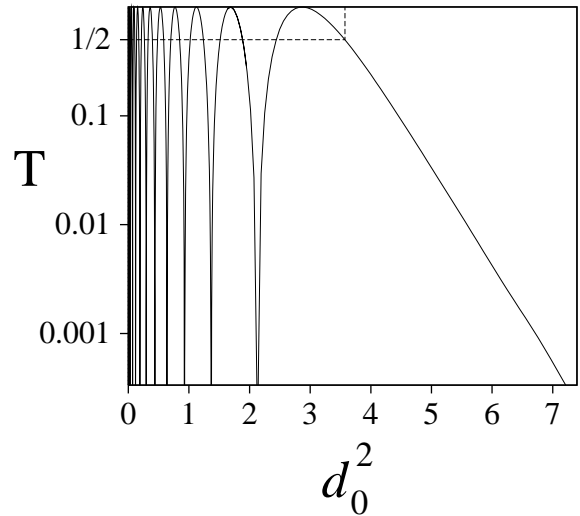}}
\caption{Tunneling probability as a function of the square of the
minimum distance between the valleys, $d_0^2$. The point of
50-50 beam splitter operation, $d_0=1.8903$, is indicated with 
dashed lines.
For small values of $d_0^2$, the wave packet is transferred back and
forth between the wells and resonant transmission occurs. For larger 
values, $d_0^2 > 3$, the relation (\ref{a9}) is seen to hold approximately.}
\label{T}
\end{figure}

Now the integration is one-dimensional for a single particle, and it is easy
to investigate the tunneling probability as a function of the parameters. In
Fig. \ref{T} we display the transition rate $T$ as a function of the
parameter $d_0^2$,
which controls the coupling between the wells. For small values,
$d_0^2 < 3,$
we are in a coherent flipping region; the wave packet is transferred back
and forth between the wells and resonant transmission occurs. For larger
values, $d_0^2 > 3,$ the analytic estimate of a logaritmic dependence in
Eq. (\ref{a9}) is seen to hold approximately. Our calculations work at
$d_0=1.8903$, which gives $T=1/2$.

\subsection{Effects of quantum statistics }

\begin{figure}
\centerline{\epsffile{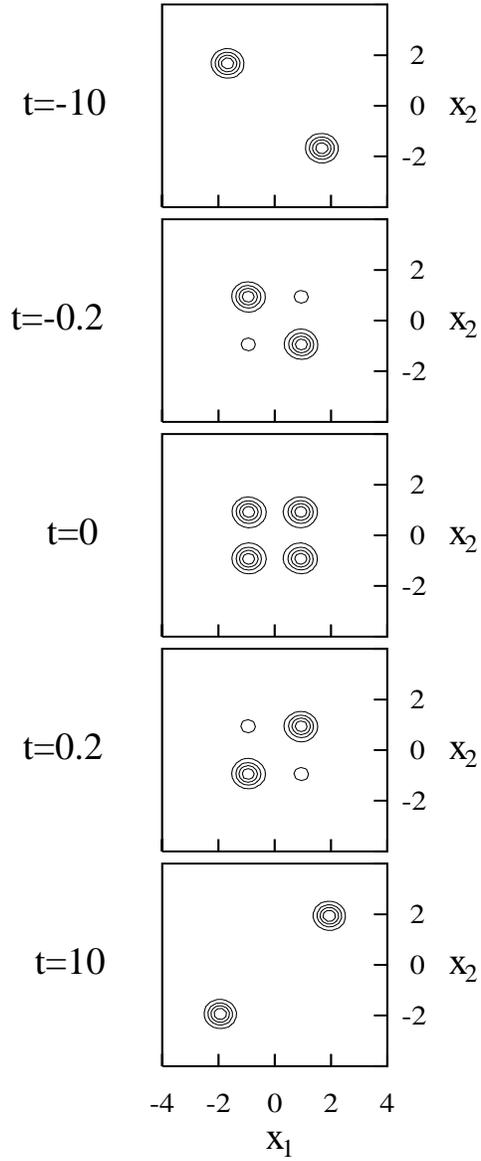}}
\caption{Two bosons propagating through the beam splitter.
Snapshots of the two-particle wave-function $|\Psi|^2$ at 
different times are shown. On the horizontal axes we see the 
coordinate of particle 1;
the vertical axes refer to particle 2. The two particles are seen
incident in different input channels; they mix around $t=0$ and
finally exit together. Parameters as in Fig. 
\ref{1dwave}.}
\label{boseV0}
\end{figure}

We can now integrate the propagation of a two-particle wave function by
choosing the initial state to be combinations of
\begin{equation}
\varphi _{L(R)}^0(x)=N\exp \left[ -{\omega\over{2\hbar}}
\left( x\pm (1+{1\over 2}d_0)\right)^2 \right] ,  \label{a31}
\end{equation}
where the $+$ $(-)$ refers to the particle entering in the left (right)
channel. For bosons, this is used in the combination (\ref{a15}) and
integrated in the potential (\ref{a7}), where the $z-$dependence is replaced
by a $t-$dependence according to Eq. (\ref{a30}). The result is shown in
Fig. \ref{boseV0}. At $t=-10$, the bosons enter symmetrically in the 
two input channels,
i.e. they have different signs for their coordinates. After being mixed at
time $t=0$, they emerge together with equal strength at both output
channels, i.e. their coordinates have the same sign. This result fully
reproduces the behavior expected from bosons at a 50-50 beam splitter.

\begin{figure}
\centerline{\epsffile{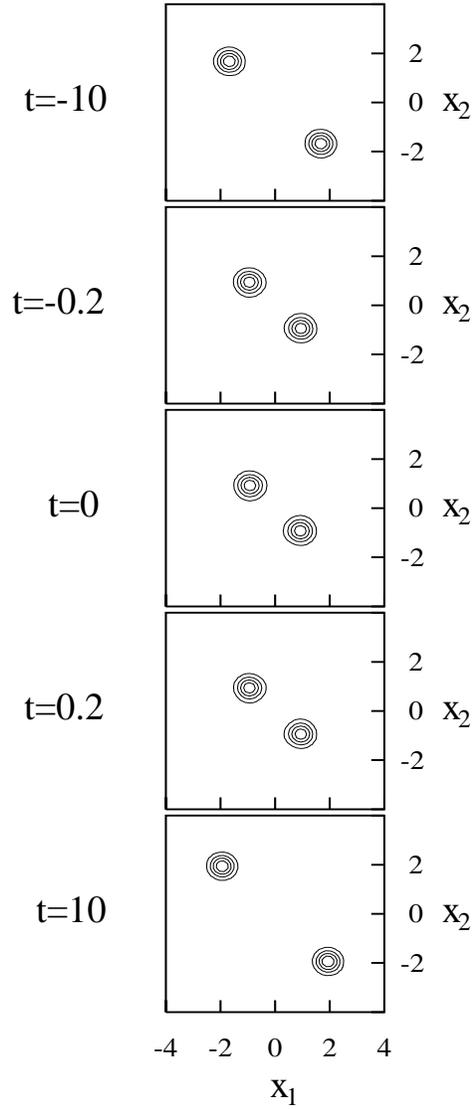}}
\caption{Two fermions propagating through the beam splitter. Snapshots
 of the two-particle wave function $|\Psi|^2$ at different times are shown.
On the horizontal axes we see the coordinate of particle 1; the 
vertical axes refer to particle 2. The two particles enter and exit
in different input and output channels. Parameters as in Fig. \ref{1dwave}.}
\label{fermiV0}
\end{figure}

We can, however, also test the fermionic case by using the state (\ref{a18})
as the initial one. The result is shown in Fig. \ref{fermiV0}. 
Near $t=0$, the wave
packets follow the potential wells, but they remain separated and emerge at
different outputs as expected. Fermions do not like to travel together.

We have thus been able to verify the properties of a 50-50 beam splitter on
massive particles represented by wave packets travelling in potential
structures. The calculations in Figs. (\ref{boseV0}) and (\ref{fermiV0}) 
do not, however, include any
particle interactions. We can now proceed to include these, and evaluate
their effect on the manifestations of quantum statistics.

When we introduce the interaction, we have to decide which type of physical
system we have in mind. Conduction electrons or ions interact through the
Coulomb force whereas neutral atom interactions may be described by a force
of the Lennard-Jones form. In both cases, the interaction is singular at the
origin, and it has to be regularized there. We do this by introducing the
variable 
\begin{equation}
r_\varepsilon =\sqrt{r^2+\varepsilon ^2}.  \label{a32}
\end{equation}
This makes the interaction energy finite when the particles overlap, 
but does not affect the main part of our argument in other ways.

With this notation the Coulomb interaction is written 
\begin{equation}
V_C(r)=\frac{V_0}{r_\varepsilon },  \label{a33}
\end{equation}
and the Lennard-Jones interaction is 
\begin{equation}
V_{LJ}(r)=V_0\left[ \left( \frac b{r_\varepsilon }\right) ^{12}-\left( \frac
b{r_\varepsilon }\right) ^6\right] ;  \label{a34}
\end{equation}
this contains the additional range parameter $b$. In both cases, the strength
of the interaction is regulated by $V_0.$

It is straightforward to integrate the Schr\"{o}dinger equation with the
two-particle interaction included, and look how its increase affects the
correlations imposed by quantum statistics. For fermions, the effect is
essentially not seen in the parameter ranges we are able to cover. As seen
from Fig. \ref{fermiV0}, the particles never really approach each other, 
and they
remain separated due to their quantum statistics for all times; the
interaction does not affect them.

\begin{figure}
\centerline{\epsffile{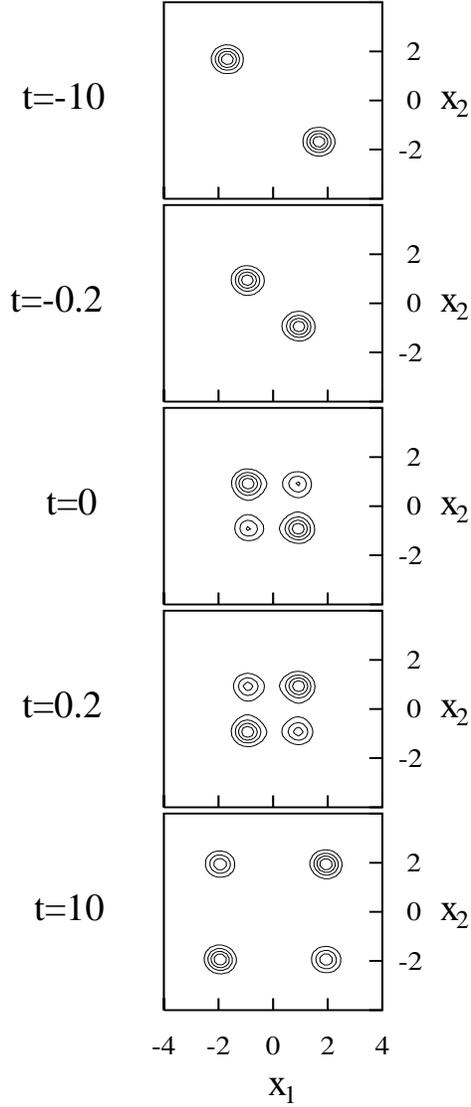}}
\caption{Two bosons propagating through the beam splitter; interaction
is included. 
Snapshots of the two-particle wave function $|\Psi|^2$ at different
times are shown. On the horizontal axes we see the coordinate of
particle 1; the vertical axes refer to particle 2. The two particles
enter in different input channels and mix around $t=0$. With no 
interaction the two bosons were always emerging together, as in Fig.
\ref{boseV0}. Now there is a finite probability also for the bosons to
emerge in separate output channels. The interaction is given by Eq. 
(\ref{a33}),  with $V_0=50$ and $\varepsilon=1$. Other parameters as in 
Fig. \ref{1dwave}.}  
\label{boseV50}
\end{figure}

\begin{figure}
\centerline{\epsffile{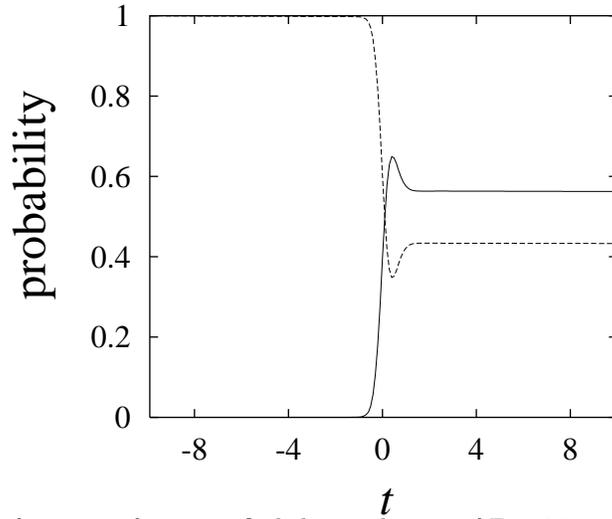}}
\caption{The probabilities, as functions of time, to find the two 
bosons of Fig. \ref{boseV50} in different valleys (full line) and in
the same valley (dashed line).}
\label{probV50}
\end{figure}

For bosons the effect is different. We have investigated their behavior for
a range of interaction parameters and find that an increase in the
interaction does destroy the simple behaviors found for noninteracting ones.
The result of such an integration is found in
Fig. \ref{boseV50}. Compared with Fig. \ref{boseV0}
this shows that now the particles appear in separate channels with about the
same probability as in the same channels. Thus the statistical effect has
been destroyed. Fig. \ref{probV50} shows how the result emerges during the time
evolution; the system tries to achieve the ideal case, but settles to the
final state observed in Fig. \ref{boseV50}.

\begin{figure}
\centerline{\epsffile{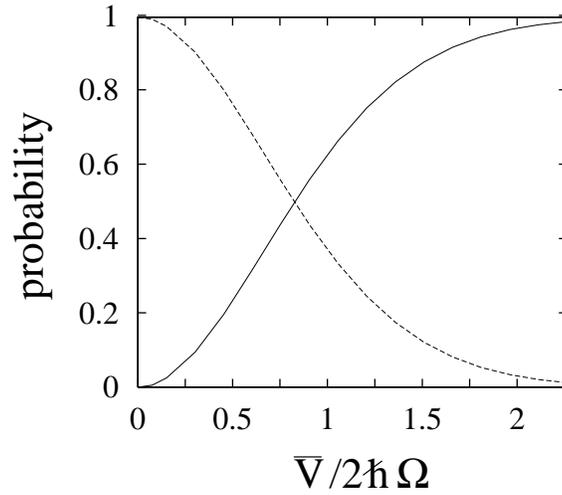}}
\caption{The probabilities, as functions of interaction strength, for two
bosons to emerge at different (full line) and same (dashed line)
output channels. The Lennard-Jones interaction is given by 
Eq. (\ref{a34}), with $b=0.25$
and $\varepsilon=0.2$. Parameters for the potential are as in Fig. 
\ref{1dwave}.}
\label{ljint}
\end{figure}

Figure \ref{ljint} shows how an increase in the interaction strength 
destroys the ideal
behavior. This is drawn using a Lennard-Jones potential, but the behavior
is similar for other cases we have investigated.

One unexpected feature emerged from our calculations: The destruction of the
ideal behavior turned out to be independent of the sign of the interaction.
One may have expected an attractive interaction between the bosons to
enhance their tendency to appear together, but this turned out not to be the
case. In order to understand this feature we have to turn to our simple
analytic argument in Sec. \ref{smod} and investigate the interplay between 
two-particle states and the interaction.

\section{Statistics versus interactions}

In order to introduce the particle interaction, we choose a convenient basis
for the two-particle states. As the first component we choose the bosonic
state (\ref{a15}) 
\begin{equation}
u_1=\frac 1{\sqrt{2}}\left( \varphi _L(1)\varphi _R(2)+\varphi _L(2)
\varphi_R(1)\right) .  \label{a35}
\end{equation}
In addition, there are two more bosonic states, where the particles enter in
the same channels 
\begin{eqnarray}
u_2 & = & \frac 1{\sqrt{2}}\left( \varphi _L(1)\varphi _L(2)+\varphi
_R(2)\varphi _R(1)\right)  \\ 
& = & \frac 1{\sqrt{2}}\left( \psi _A(1)\psi _A(2)+\psi _S(1)\psi
_S(2)\right)
\label{a36}
\end{eqnarray}
and 
\begin{eqnarray}
u_3 & = & \frac 1{\sqrt{2}}\left( \varphi _L(1)\varphi _L(2)-\varphi
_R(2)\varphi _R(1)\right)\label{a37}  \\ 
& = & \frac 1{\sqrt{2}}\left( \psi _S(1)\psi _A(2)+\psi _A(1)\psi
_S(2)\right) .\nonumber
\end{eqnarray}
As the last component we have the fermionic basis function (\ref{a18}) 
\begin{equation}
u_4=\frac 1{\sqrt{2}}\left( \varphi _L(1)\varphi _R(2)-\varphi _L(2)\varphi
_R(1)\right) .  \label{a38}
\end{equation}
Together the functions $\{u_i\}$ form a complete Bell state basis for the
problem. They are also convenient for the introduction of particle
interactions. In the states $u_1$ and $u_4$ the wave functions overlap only
little, and the effect of the interaction is small. For the states 
$u_2$ and $u_3$ they sit on top of each other and feel the interaction
strongly. We treat this in a Hubbard-like fashion by saying that the energy
of the latter states is changed by the value 2$\overline{V}\propto V_0$.
Because the overlap between $\varphi_L$ and $\varphi_R$ is small, we have
$(u_1,Vu_1)\approx (u_4,Vu_4)$ and $(u_2,Vu_2)\approx (u_3,Vu_3)$, and
we can use the definition
\begin{eqnarray}
2\overline{V}&=&{1\over 2}\left[(u_2,Vu_2)+(u_3,Vu_3)-
(u_1,Vu_1)-(u_4,Vu_4)\right]\\
&\approx&\int\int\varphi(x)^2\varphi(y)^2 V(|x-y|) dx dy- 
\int\int\varphi(x)^2\varphi(y-d_0)^2 V(|x-y|) dx dy, \nonumber
\end{eqnarray} 
where $V$ is either a Coulomb or Lennard-Jones interaction. 
The first terms give the effective interaction energy when both
particles sit in the same potential groove. In the states $u_1$ and
$u_4$ both grooves are occupied and because they are at their closest
at $z=0$, we subtract the mutual interaction energy across the 
separating barrier to obtain the pure local interaction energy. 
The second line results if we approximate both $\varphi_L$ and 
$\varphi_R$ by a Gaussian $\varphi$ with the same width. From
Eq. (\ref{a17}) we see that only the states $u_1$ and $u_2$ are coupled by
the tunneling rate $2\Omega .$ Thus if we express the state by 
\begin{equation}
\Psi =a_1u_1+a_2u_2+a_3u_3+a_4u_4,  \label{a39}
\end{equation}
the state vector $[a_1,a_2,a_3,a_4]$ evolves with the Hamiltonian 
\begin{equation}
\left[ 
\begin{array}{cccc}
2\overline{E} & -2\hbar\Omega  & 0 & 0 \\ 
-2\hbar\Omega  & 2\overline{E}+2\overline{V} & 0 & 0 \\ 
0 & 0 & 2\overline{E}+2\overline{V} & 0 \\ 
0 & 0 & 0 & 2\overline{E}
\end{array}
\right] =2 \overline{E}+\overline{V} +\left[ 
\begin{array}{cccc}
-\overline{V} & -2\hbar\Omega & 0 & 0 \\ 
-2\hbar\Omega  & \overline{V} & 0 & 0 \\ 
0 & 0 & \overline{V} & 0 \\ 
0 & 0 & 0 & -\overline{V}
\end{array}
\right] .  \label{a40}
\end{equation}
The constant part does not affect the coupling between the states, and the
amplitudes $a_3$ and $a_4$ decouple. The remaining ones flip at the
effective rate 
\begin{equation}
\Omega _{\rm{eff}}=\frac 12\sqrt{4\Omega^2 
+\overline{V}^2/\hbar^2 }.  \label{a41}
\end{equation}
This result shows that the new parameter replacing $\Omega$ in
Eq. (\ref{a11}) is $\Omega
_{\rm{eff}},$ implying that the perfect boson behavior is expected to be
destroyed for 
\begin{equation}
\frac{\overline{V}}{2\hbar\Omega }\sim \sqrt{3}=1.73.  \label{a42}
\end{equation}
This is in approximate agreement with the result shown in Fig. \ref{ljint}.
By inspecting Fig. \ref{probV50}, we can also verify that the flipping
does occur
faster when we switch on the interaction, as expected from Eq. (\ref{a41}).

In the simplified analytic treatment, the only influence of the potential
was through its strength $| \overline{V}| .$ Hence we expect the
results to scale with the parameter 
$\left( | \overline{V}| /2\hbar\Omega
\right) $ where $2\hbar\Omega =E_A-E_S.$ 
The probability to emerge in the same
output channels should essentially depend on this only; a certain
universality is expected.

\begin{figure}
\centerline{\epsffile{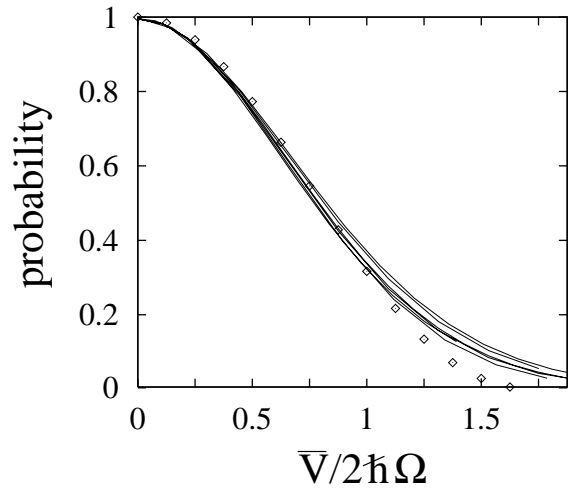}}
\caption{The probability, as a function of interaction strength, for two
bosons to emerge at the same output channel  
for several different types of interaction (full lines).
We see that the perfect bosonic behavior is destroyed for 
$\overline{V}/2\hbar\Omega \sim \sqrt{3}=1.73$. 
The parameter ranges are $\varepsilon=0.1$ to 1 for the Coulomb
interaction and $b=0.25$ to 0.5, $\varepsilon=0.2$ to 0.35 for the
Lennard-Jones interaction. 
The numerical results are compared with that of the analytic 
treatment (diamonds). In this model $2\hbar\Omega$ was chosen to be 8, and 
the time evolution in the subspace $\{ u_1, u_2 \} $ was calculated.}
\label{univ}
\end{figure}

In Fig. \ref{univ}, we have plotted this probability for a variety 
of potentials including both Coulomb and Lennard-Jones ones. As we 
can see, the behavior is very similar, at 
$\left( | \overline{V}| /2\hbar\Omega \right)\approx 1.7$ 
the probability has decreased to less than 10\% in agreement with our
expectation. This verifies the degree of universality achieved. 
For comparison, we also used the simple analytic theory to obtain the
points along the curve.
In this treatment, $2\Omega$ was assumed to be constant during
some finite coupling time $t$, according to Eq. (\ref{a14}). By inspecting
Fig. \ref{probsplit}, we conclude that $t$ should be of the order of unity. 
Here $2\hbar\Omega$ was chosen to be 8, and the time
evolution in the subspace $\{ u_1, u_2 \} $ was calculated. This
agrees best with the numerical results for small $\overline{V}$;  for larger
values of $\overline{V}$ the simple analytic treatment becomes invalid.

\section{Discussion and Conclusions}

The actual values of $\xi$, $\tau$, $m$ and $\omega$ depend on the 
physical system  we have in mind. Our calculations are carried out at
$\widetilde{\omega}=30$ and $\widetilde{\hbar}=6$; the momentum in the
$z$-direction of the incoming wave packet is $\widetilde{p}_z=30$ or
$\widetilde{p}_z=1000$. 
If we consider an atomic beam splitter for Rb atoms,
setting the length scale $\xi$ to 100~nm corresponds to a time scale
$\tau$ of $80~\mu$s according to Eq. (\ref{a23}). A displacement of 
400~nm from the center of one valley in the $x$-direction gives 
a potential energy increase of roughly 10~mK, i.e. a transverse
velocity of 0.15~m/s. This is to be compared with a typical height of 
the confining potential in a hollow optical fiber, a few tens of 
mK \cite{ito,yin}. Taking $\widetilde{p}_z=1000$ yields a beam
velocity of 1.3~m/s in the $z$-direction.

If we consider a mesoscopic electron beam splitter built on GaAs, 
$\xi$ could be of the order of 40 nm, which means that the closest 
distance $d_0$ between the valleys is 80 nm. This would correspond to
$\tau=6$ ps, i.e. the electron goes through the device in a few
picoseconds. With $\widetilde{p}_z=30$, the
kinetic energy of the electron due to the
motion in  the $z$-direction would be of the order of 0.01~eV.
A displacement of 100 nm from the 
center of one valley in the $x$-direction corresponds to a potential 
energy increase of roughly 0.05~eV, in comparison with the
bandgap in GaAs, 0.115~eV.

The parameter ranges chosen in our illustrative computations may not
be experimentally optimal, but they indicate that the effects are not
totally outside the range of real systems. Even if our calculations
are based on a rather simplified model, we find them suggesting
effects possible in realistic setups. The main problem is to
prepare the appropriate quantum states, launch them into the structures
and retain their quantum coherence during the interaction. With atomic
cooling and trapping techniques, this may be feasible in the near
future. For electrons the possibility to retain quantum coherence is
still an open question.

We have chosen to discuss the straightforward question of particle 
statistics
at a beam splitter. This is a simple situation, which, however,
presents genuine quantum features. For information processing and
quantum logic slightly more complicated networks are needed. Simple
gate operation can be achieved along the lines described in
Ref. \cite{kir}, which was formulated in terms of conduction
electrons, but similar situations can be envisaged for interacting
atoms.

In this paper we suggest an analytic model, which can be used to analyze
the behavior of particle networks confined to potential grooves in
two-dimensional structures as discussed also by Schmiedmayer
\cite{sch}. For complicated situations, the full numerical treatment
rapidly becomes intractable and simplified qualitative tools are
needed. We hope to have contributed to the development of such methods
in the present paper.

\acknowledgments

One of us (SS) thanks Dr. J\"org Schmiedmayer for inspiring 
discussions. MTF thanks Patrick Bardroff for fruitful discussions.

\vfill\break


\end{document}